\documentclass[twocolumn,aps,superscriptaddress,longbibliography]{revtex4-1}
\usepackage{bm}
\usepackage{amsmath}
\usepackage{amssymb}
\usepackage{times}
\usepackage[usenames,dvipsnames]{xcolor}
\usepackage{graphicx}
\usepackage{dcolumn}
\usepackage{simplewick}
\usepackage{hyperref}

\hypersetup{
  colorlinks=false,
  colorlinks=true,
  citecolor=blue,
  linkcolor=blue,
  urlcolor=blue}

\DeclareUnicodeCharacter{2212}{-}
\begin{document}

\title{Fock-space relativistic coupled-cluster calculations of clock 
       transition properties in Pb$^{2+}$}

\author{Palki Gakkhar}
\affiliation{Department of Physics, Indian Institute of Technology,
             Hauz Khas, New Delhi 110016, India}

\author{Ravi Kumar}
\affiliation{Department of Chemistry, University of Zurich,
             Switzerland}             

%\author{Suraj Pandey}
%\affiliation{Department of Physics, Indian Institute of Technology,
%             Hauz Khas, New Delhi 110016, India}

%%%%%%%%%%%
\author{D. Angom}
\affiliation{Department of Physics, Manipur University,
             Canchipur 795003, Manipur, India}

\author{B. K. Mani}
\email{bkmani@physics.iitd.ac.in}
\affiliation{Department of Physics, Indian Institute of Technology,
             Hauz Khas, New Delhi 110016, India}             

\begin{abstract}

We have implemented an all-particle multireference Fock-space relativistic 
coupled-cluster theory to probe $6s^2{\;^1}S_{0} - 6s6p{\;^3P^o_{0}}$ clock 
transition in an even isotope of Pb$^{2+}$. We have computed, 
excitation energy for several low lying states, E1 and M1 transition amplitudes, and 
the lifetime of the clock state. Moreover, we have also calculated the ground 
state dipole polarizability using perturbed relativistic coupled-cluster theory. 
To improve the accuracy of results, we incorporated the corrections from 
the relativistic and QED effects in all our calculations. The contributions from
triple excitations are accounted perturbatively. Our computed excitation energies 
are in excellent agreement with the experimental values for all the states. 
Our result for lifetime, $9.76\times10^{6}$ s, of clock state is $\approx$ 8.5\% 
larger than the previous value using 
CI+MBPT [Phys. Rev. Lett. {\bf 127}, 013201 (2021)].
Based on our analysis, we find that the contributions from the {\em valence-valence}
correlations arising from higher energy configurations and the corrections from the 
perturbative triples and QED effects are essential to get accurate clock 
transition properties in Pb$^{2+}$. Our computed value of dipole polarizability 
is in good agreement with the available theoretical and experimental data.

\end{abstract}

%% 31.15.bw: Coupled-cluster theory
%% 11.30.Er: Charge conjugation, parity, time reversal, and other 
%%           discrete symmetries 
%% 31.15.am: Relativistic configuration interaction (CI) and many-body 
%%           perturbation calculations

\maketitle

%%%%%%%%%%%%%%%%%%%%%%%%%%%%%%%%%%%%%%%%%%%%%%%%%%%%%%%%%%%%%%%%%%%%%%%
%%%%                          Introduction                         %%%%
%%%%%%%%%%%%%%%%%%%%%%%%%%%%%%%%%%%%%%%%%%%%%%%%%%%%%%%%%%%%%%%%%%%%%%%
\section{Introduction}

Optical atomic clocks are one of the most accurate time measurement instruments 
in existence today \cite{ludlow-15, subhadeep-23}. Due to their unprecedented 
accuracies as frequency and time standards, they can 
serve as important probes of fundamental phenomena in physics and 
function as key components in technological applications. 
Some examples where atomic clocks are of vital importance include, 
measuring the variation in the fundamental constants \cite{safronova-19, 
prestage-95, karshenboim-10}, probing physics beyond the standard model 
of particle physics \cite{dzuba-18, berengut-18},
navigation systems \cite{grewal-13, major-13}, 
quantum computer \cite{ weiss-17, wineland-02}, the basis for redefining 
the second \cite{karshenboim-10,riehle-18}, 
and others \cite{ludlow-15, subhadeep-23}. For the single ion optical clocks, 
the hyperfine induced $3s^2{\;^1}S_0-3s3p{\;^3}P^o_0$ 
(267.4 nm) transition based ${^{27}}$Al$^+$ is demonstrated to be one of 
the best clocks, with a fractional frequency uncertainty of 
$9.4\times10^{-19}$ \cite{brewer_19a}. The high accuracy 
in ${^{27}}$Al$^+$ could be attributed to the low sensitivity to 
electromagnetic fields, narrow natural linewidth and small room temperature 
black-body radiation (BBR) shift in the clock transition 
frequency \cite{kallay-11,safronova-11, ravi-21a}.  
Among the neutral atoms, a lattice clock based on degenerate fermionic $^{87}$Sr 
atoms with a hyperfine-induced $5s^2{\;^1}S_0 - 5s5p{\;^3}P^o_0$ (698 nm) 
transition is reported to be one of the best neutral atom clocks. The smallest 
fractional frequency error achieved is $\approx$ 2.0 $\times$ 10$^{-18}$ 
\cite{nicholson-15, bothwell-19}.

In the quest for a new and improved frequency standard, 
an optical clock based on 6s$^{2}{\;^1S_{0}} - 6s6p{\;^3}P^o_{0}$ transition, 
mediated through a {\em two} photon E1+M1 channel, in doubly ionized even 
isotope of Lead (Pb$^{2+}$) could be a promising candidate.
Like in ${^{27}}$Al$^+$, the clock transition is an electric dipole 
forbidden transition between two $J=0$ states, providing a strong resistance 
to the environmental perturbations. In addition, unlike ${^{27}}$Al$^+$, the 
nuclear spin quantum number $I$ is zero. This is crucial, as it prevents clock 
transition from the nonscalar perturbations which may arise through the coupling 
between the electron and nuclear multipole moments. Despite this important 
prospect with Pb$^{2+}$ as an accurate optical atomic clock, the properties of
the relevant transition have not been explored in detail. For example, in terms
of theoretical calculations, to the best of our knowledge, there is only one
study on the lifetime of the clock state \cite{beloy-21}. The work 
\cite{beloy-21}, employing a combined method of configuration interaction (CI) 
and many-body perturbation theory (MBPT), computed the lifetime, $\tau$, 
of the clock 
state, ${\;^3}P^{o}_{0}$, as $9.0\times10^{6}$ s. Considering that there are 
no experimental data, additional theoretical calculations, specially 
using the accurate methods like relativistic coupled-cluster (RCC), would be 
crucial to get better and accurate insights on the clock properties. Moreover, 
the inclusion of relativistic and QED corrections to the properties calculations 
are essential to obtain reliable results. It can thus be concluded that there 
is a clear research gap in terms of the scarcity of accurate properties results 
for $^1S_0 - {^3}P^o_0$ clock transition of Pb$^{2+}$.

In this work, we have implemented an all-particle multireference Fock-space 
relativistic coupled-cluster (FSRCC) theory to compute the clock transition
properties of Pb$^{2+}$ accurately. It is to be noted that, RCC theory is one 
of the most reliable quantum many-body theories for atomic structure 
calculations. It accounts for electron correlation effects to all-orders of 
residual Coulomb interaction and has been employed to obtain accurate 
results in several closed-shell and one-valence atoms and ions 
\cite{pal-07,mani-09,nataraj-11,ravi-20}. The application of RCC for 
two-valence atomic systems, as the present case of Pb$^{2+}$ clock transition 
is, however, limited to few studies \cite{eliav-95,eliav-95b,mani-11b}. And, 
the reason for this, perhaps, is the complications associated with the 
implementation of FSRCC theory for 
multi-reference systems \cite{eliav-95,eliav-95b,mani-11b,ravi-21a}.
To address the clock transition properties in a comprehensive way, using 
FSRCC theory \cite{mani-11b, ravi-21a}, we carried out precise calculations 
of the excitation energies 
and E1 and M1 transition amplitudes associated with $^1S_0 - {^3}P^o_0$ 
transition in Pb$^{2+}$. Using these results, we have then calculated 
the lifetime of the $^3P^o_0$ clock state. In addition, as electric dipole 
polarizability is a crucial parameter for estimating BBR shift in clock 
frequency, we have also calculated the ground state polarizability of Pb$^{2+}$ 
using perturbed relativistic coupled-cluster (PRCC) theory 
\cite{ravi-20, ravi-21b}. Moreover, in all these properties calculations, 
we have incorporated and analyzed the contributions from the Breit interaction, 
QED corrections and perturbative triples.

The remainder of the paper is organized into five sections. In Sec. II, we 
provide  a brief description of the FSRCC theory for two-valence atomic systems. 
We have given the coupled-cluster working equation for two-valence systems. In 
the Sec. III, we provide and discuss the expression for E1M1 decay rate.  The 
results obtained from our calculations are presented and analyzed in Sec. IV. 
Theoretical uncertainty in our computed results is discussed in Sec. V of the 
paper. Unless stated otherwise, all results and equations presented in this 
paper are in atomic units ( $\hbar=m_e=e=1/4\pi\epsilon_0=1$).

%%%%%%%%%%%%%%%%%%%%%%%%%%%%%%%%%%%%%%%%%%%%%%%%%%%%%%%%%%%%%%%%%%%%%%%%%%%%%%
%%%%%               The theoretical methods                             %%%%%%
%%%%%%%%%%%%%%%%%%%%%%%%%%%%%%%%%%%%%%%%%%%%%%%%%%%%%%%%%%%%%%%%%%%%%%%%%%%%%%

\section{Two-valence FSRCC Theory}

Since the clock transition in Pb$^{2+}$ involves atomic state functions (ASFs) 
of two-valence nature, we need an accurate multireference theory to calculate 
these wavefunctions and corresponding many-body energies. In the present work 
we have employed a FSRCC theory for two-valence \cite{mani-11b, ravi-21a} 
to obtain the many-body wavefunction and corresponding energy. 
In Refs. \cite{mani-11b, mani-17, ravi-21a}, we have discussed in detail 
the implementation of FSRCC theory to sophisticated parallel codes and 
have also given the working 
equations and Goldstone diagrams contributing to the theory. So, here, 
for completeness, we provide a very brief description of FSRCC theory for 
two-valence atoms and properties calculations using it in the 
context of Pb$^{2+}$.

\begin{figure}
 \includegraphics[scale=0.6, angle=0]{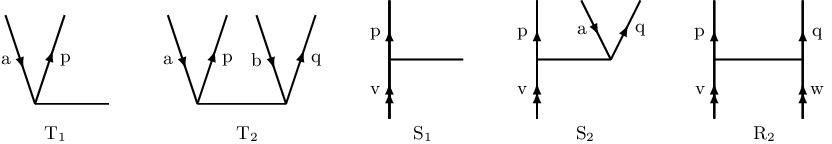}
	\caption{The diagrammatic representation of closed-shell, one-valence, 
	and two-valence single and double CC operators.}
 \label{tsr-diag}
\end{figure}

The atomic state function for a two-valence atom or ion is obtained 
by solving the many-body Schrodinger equation
\begin{equation}
  H^{\rm DCB}|\Psi_{vw} \rangle = E_{vw} |\Psi_{vw} \rangle,
  \label{hdc_2v}
\end{equation}
where $|\Psi_{vw}\rangle$ is the exact many-body wavefunction and $E_{vw}$ 
is the corresponding exact energy. The indices $v, w, \cdots$ represent the 
valence orbitals. $H^{\rm DCB}$ is the Dirac-Coulomb-Breit no-virtual-pair 
Hamiltonian used in all calculations, and expressed as 
\begin{eqnarray}
   H^{\rm DCB} & = & \sum_{i=1}^N \left [c\bm{\alpha}_i \cdot
        \mathbf{p}_i + (\beta_i -1)c^2 - V_{N}(r_i) \right ]
                       \nonumber \\
    & & + \sum_{i<j}\left [ \frac{1}{r_{ij}}  + g^{\rm B}(r_{ij}) \right ],
  \label{ham_dcb}
\end{eqnarray}
where $\bm{\alpha}$ and $\beta$ are the Dirac matrices, and $1/r_{ij} $ 
and $g^{\rm B}(r_{ij})$ are the Coulomb and Breit interactions, respectively.
In FSRCC theory, $|\Psi_{vw} \rangle$ is written as
\begin{equation}
|\Psi_{vw}\rangle = e^T \left[ 1 + S_1 + S_2 + 
	\frac{1}{2} \left({S_1}^2 + {S_2}^2 \right) + 
	R_2 \right ]|\Phi_{vw}\rangle,
  \label{2v_exact}
\end{equation}
where $|\Phi_{vw}\rangle, = a^\dagger_wa^\dagger_v |\Phi_0\rangle,$
is the Dirac-Fock reference state for a two-valence system. Operators 
$T$, $S$ and $R$ are the electron excitation operators, referred to as 
the coupled-cluster (CC) operators, for closed-shell, one-valence 
and two-valence sectors, respectively. The subscripts $1$ and $2$ with 
these operators represent the single and double excitations, referred 
to as the coupled-cluster with singles and doubles (CCSD) approximation.
The FSRCC theory with CCSD approximation subsumes most of the electron 
correlation effects in atomic structure calculations and provides an 
accurate description of the calculated properties. In the second quantized 
representation, the CC operators are expressed as
\begin{subequations}
\begin{eqnarray}
   T_1  = \sum_{ap}t_a^p a_p^{\dagger}a_a {\;\; \rm and \;\;} 
   T_2  = \frac{1}{2!}\sum_{abpq}t_{ab}^{pq} a_p^{\dagger}a_q^{\dagger}a_ba_a,
\end{eqnarray}
\begin{eqnarray}
   S_1 = \sum_{p}s_v^p a_p^{\dagger}a_v  {\;\; \rm and \;\;}
   S_2 = \sum_{apq}s_{va}^{pq} a_p^{\dagger}a_q^{\dagger}a_aa_v,
\end{eqnarray}
\begin{eqnarray}
  R_2 = \sum_{pq}r_{vw}^{pq} a_p^{\dagger}a_q^{\dagger}a_wa_v.
\end{eqnarray}
 \label{t1t2}
\end{subequations}
Here, the indices $a, b,\ldots$ and $p, q,\ldots$ represent the core and virtual 
orbitals, respectively. And, $t_{\ldots}^{\ldots}$, $s_{\ldots}^{\ldots}$ 
and $r_{\ldots}^{\ldots}$ are the cluster amplitudes corresponding to $T$, 
$S$ and $R$ coupled-cluster operators, respectively. The diagrammatic 
representation of these operators is shown in Fig. \ref{tsr-diag}. 
It is to be however mentioned that, the dominant contributions from triple 
excitations are also included using perturbative triples approach \cite{ravi-21a}. 

%%%%%%%%%%%%%%%% figure 1

The operators for closed-shell and one-valence sectors are obtained 
by solving the set of coupled nonlinear equations discussed in 
Refs. \cite{mani-09} and \cite{mani-10}, respectively. 
%The details related to the computational implementation of RCC method 
%for closed-shell and one-valence systems are given in Ref. \cite{mani-17}, 
%where have published our code. 
The two-valence CC operator, $R_2$, is obtained by solving 
the CC equation \cite{mani-11b, ravi-21a}
\begin{eqnarray}
   \langle\Phi^{pq}_{vw}|
    \bar H_{\rm N} +
   \{\contraction{\bar}{H}{_{\rm N}}{S}\bar H_{\rm N}S^{'}\} +
   \{\contraction{\bar}{H}{_{\rm N}}{S}\bar H_{\rm N}R_2\}
   |\Phi_{vw}\rangle = \nonumber \\
   E^{\rm att}_{vw}
   \langle\Phi^{pq}_{vw}|\Bigl[S^{'} + R_2 \Bigr]|\Phi_{vw}\rangle.
   \label{ccsd_2v}
\end{eqnarray}
Here, for compact notation we have used $S'=S_1+S_2 + \frac{1}{2}(S^2_1 + S^2_2)$. 
$E^{\rm att}_{vw}$ is the two-electron attachment energy and it is expressed 
as the difference between the correlated energy of $(n-2)-$electron (closed-shell)
and $n-$electron (two-valence) sectors, $E_{vw} - E_0$. 

%%%%%%%%%%%%%%%% figure 1
\begin{figure}
 \includegraphics[scale=0.5, angle=0]{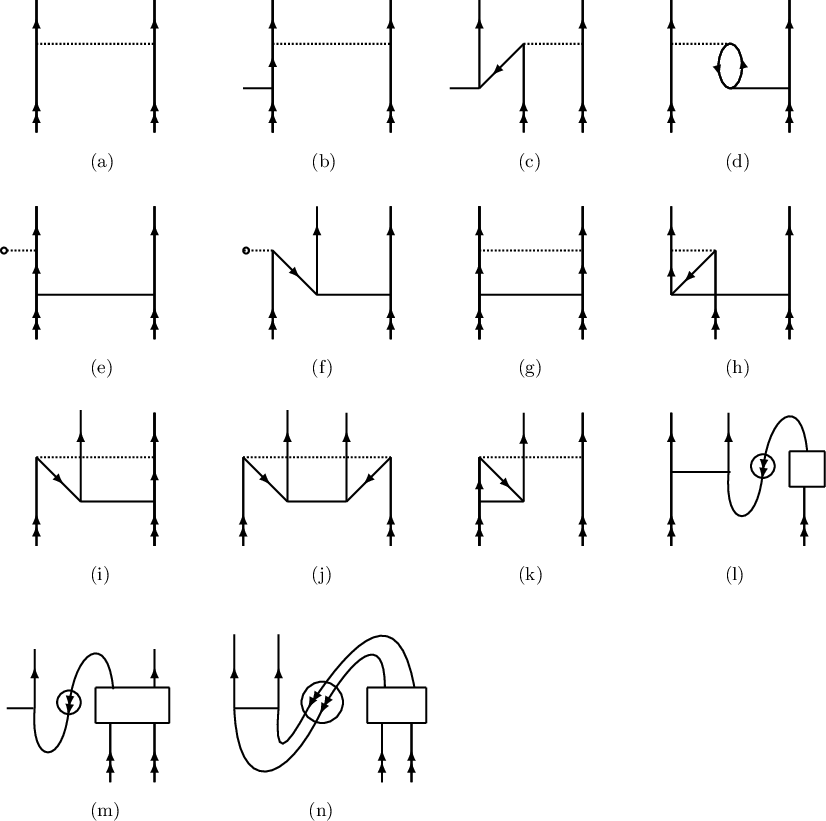}
	\caption{The CC diagrams contributing to the {\em linearized} FSRCC
	theory for two-valence atomic systems. Diagrams (l) - (n) are referred 
	to as the folded diagrams and arise from the renormalization terms 
	in CC equations for multireference systems.}
 \label{2v-diag}
\end{figure}

In Fig. \ref{2v-diag}, we have given the Goldstone diagrams contributing 
to {\em linearized} FSRCC theory for two-valence systems. These are 
obtained by considering the terms with only {\em one} order of CC 
operators in Eq. (\ref{ccsd_2v}) and then contracting residual Coulomb 
interaction with these CC operators using Wick's theorem. The CC diagrams 
$(l) - (n)$ are referred to as the folded diagrams and they arise due 
to renormalization terms on the right hand side of Eq. (\ref{ccsd_2v}). 
The presence of folded diagrams in open-shell systems constitute one of
the main differences from the CC theory of closed-shell systems. The 
rectangular portion represents the effective energy diagrams arising from 
the one-valence (diagram (l)) and two-valence (diagrams (m) and (n)) sectors.
The Goldstone diagrams in Fig. \ref{2v-diag} correspond to the algebraic 
expression 
\begin{eqnarray}
	\langle & H_{\rm N} & \rangle^{pq}_{vw} + 
   \langle \contraction{}{H}{_{\rm N}}{T} H_{\rm N}T \rangle^{pq}_{vw} +
   \langle \contraction{}{H}{_{\rm N}}{S} H_{\rm N}S^{'} \rangle^{pq}_{vw} +
   \langle \contraction{}{H}{_{\rm N}}{S} H_{\rm N}R_2 \rangle^{pq}_{vw} - \nonumber \\
   && \langle \contraction{}{E}{^{\rm att}_{vw}}{E} E^{\rm att}_{vw} S^{'} \rangle^{pq}_{vw} -  
   \langle \contraction{}{E}{^{\rm att}_{vw}}{R} E^{\rm att}_{vw} R_2 \rangle^{pq}_{vw} =
	g_{pqvw} + g_{pqrw}s^r_v  \nonumber \\
    &&	- g_{aqvw}t^p_a + \tilde{g}_{pavr}s^{rq}_{aw} + g_{pqvw}\epsilon_p + g_{pqrs}r^{rs}_{vw} \nonumber \\
    && - g_{aqvr}s^{pr}_{aw} + g_{abvw}t^{pq}_{ab} - g_{aqrw}s^{rp}_{va} - E^{\rm att}_w r^{pq}_{vw} \nonumber \\
    && - E^{\rm att}_{vw}s^{p}_{v} - E^{\rm att}_{vw} r^{pq}_{vw},
\end{eqnarray}
where $\tilde{g}_{ijkl} = g_{ijkl} - g_{ijlk}$. Since we have used Dirac-Fock 
orbitals in our calculations, diagram (f) does not contribute, and therefore 
not included in the expression.

%%%%%%%%%%%%%%%%%%%%%%%%%%%%%%%%%%%%%%%%%%%%%%%%%%%%%%%%%%%%%%%%%%%%%%%%%%%%%%
%%%%%%%%%%%%%%%%%   E1M1 transition amplitude  %%%%%%%%%%%%%%%%%%%%%%%%%%%%%%%
%%%%%%%%%%%%%%%%%%%%%%%%%%%%%%%%%%%%%%%%%%%%%%%%%%%%%%%%%%%%%%%%%%%%%%%%%%%%%%
\section{E1M1 Decay Rate Using FSRCC}

\begin{figure}
 \includegraphics[scale=0.32, angle=0]{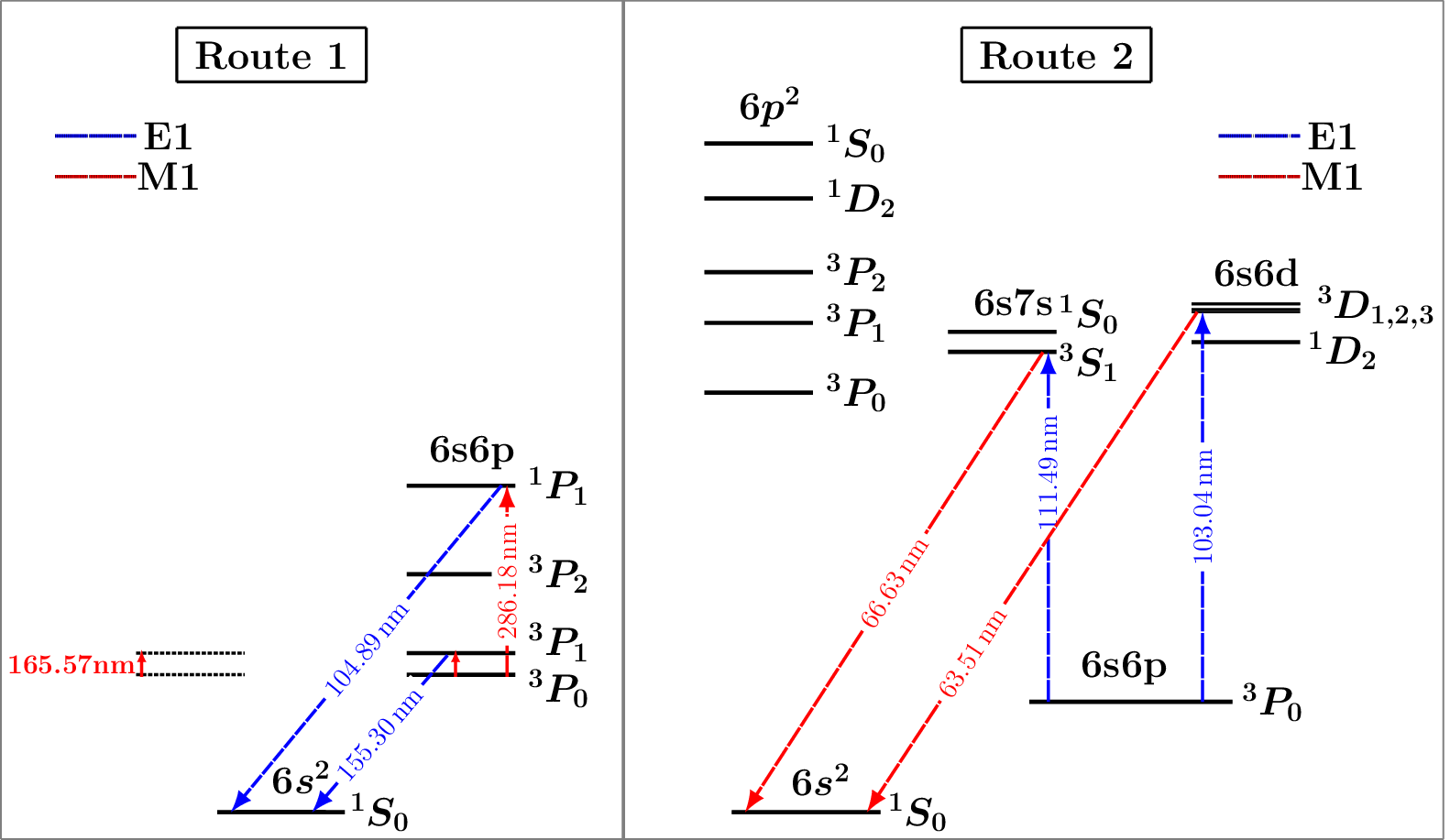}
	\caption{Schematic energy level diagram for $6s^2\;^1S_0 \rightarrow 6s6p\;^3P^{o}_0$ 
	clock transition in Pb$^{2+}$ via a {\em two} photon E1+M1 transition.}
 \label{fig_e1m1}
\end{figure}

Since $I = J = F = 0$ for $^1S_0 - {^3}P^o_0$ clock transition in Pb$^{2+}$, 
it is allowed through a {\em two-photon } E1+M1 channel. As shown 
in the schematic diagram, Fig. \ref{fig_e1m1}, in the first route, the initial 
state $|\Psi_i \rangle$ can couple to a same parity state through a magnetic 
dipole operator (photon with energy $\omega_1$) and then connect to the 
final state $|\Psi_f \rangle$ through an electric dipole operator 
(photon with energy $\omega_2$). Alternatively, in the second route, the 
initial state $|\Psi_i \rangle$ can couple to an opposite parity state via an 
electric dipole operator first and then connect to the ground state 
through a magnetic dipole operator. Mathematically, the E1+M1 decay 
rate from $|\Psi_f \rangle$ to $|\Psi_i \rangle$ can be 
expressed in terms of the reduced matrix elements of electric and 
magnetic dipole operators, as \cite{craig-98, santra-04}
\begin{eqnarray}
  \Gamma_{E1M1} = \frac{8}{27\pi}\alpha^6 && \int_{0}^{\infty}d\omega_{1}
  \omega_{1}^{3}\int_{0}^{\infty}d\omega_{2}\omega_{2}^{3} \nonumber \\
	&& \times \bigg | \frac {\langle \Psi_f||{\mathbf D}||\Psi_n\rangle \langle 
             \Psi_n||M_{1}||\Psi_i \rangle} {E_n + \omega_{1} - E_{i}}  \nonumber \\
    && + \frac {\langle \Psi_f||M_1||\bar \Psi_n\rangle \langle 
	     \bar \Psi_n||{\mathbf D}||\Psi_i \rangle} {E_{\bar n} + \omega_{2} - E_{i}}
	     \bigg |^{2} \nonumber \\
   &&   \times   \delta (E_{i} + \omega_{1} + \omega_{2} - E_{f}).
\label{e1m1}
\end{eqnarray}
Here for Pb$^{2+}$: $|\Psi_i\rangle = 6s6p{\;^3}P^{o}_0$, 
$|\Psi_f\rangle = 6s^2{\;^1}S_0$, $|\Psi_n\rangle = 6s6p{\;^3}P^{o}_1, 
6s6p{\;^1}P^{o}_1$ and $|\bar \Psi_n\rangle = 6s7s{\;^3}S_1, 6s6d{\;^3}D_1$.
Since the transition is allowed through two photons, the energy 
difference of final and initial states satisfies the relation 
$E_f - E_i = \omega_1 + \omega_2$.

The reduced matrix elements in Eq. (\ref{e1m1}) are calculated using the FSRCC 
theory. Properties calculation using FSRCC theory is explained in detail in our 
work \cite{ravi-21a}. However, to illustrate it briefly in the present work, 
we consider the example of dipole matrix element. Using the RCC wave function 
from Eq. (\ref{2v_exact}), the dipole matrix elements is
\begin{eqnarray}
	&&\langle\Psi_f|{\mathbf D}|\Psi_n\rangle = \sum_{k l}{c^f_k}^*c^n_l
	\left[\langle\Phi_k|\tilde {\mathbf D}+ \tilde {\mathbf D}
  \left(S^{'} 
               \right. \right. \nonumber \\
  &&\left.\left. + R_2 \right ) 
	+ \left(S^{'} + R_2 \right )^\dagger \tilde {\mathbf D}
    + \left(S^{'} + R_2 \right )^\dagger
               \right. \nonumber \\
	&&\left.\; \tilde {\mathbf D}\left(S^{'} + R_2 \right )
    |\Phi_l\rangle \right],
  \label{hfs_cc}
\end{eqnarray}
where the coefficients $c^f_k$ represent the mixing coefficients in 
the expansion of a multireference configuration state function 
$|\Phi_f\rangle$. These are obtained by diagonalizing the $H^{\rm DCB}$ 
matrix within the chosen model space.  The dressed operator 
$\tilde {\mathbf D},= e{^T}^\dagger {\mathbf D} e^T,$ is a non 
terminating series in closed-shell CC operator $T$. Including all 
orders of $T$ in the dressed operator is practically challenging.
In Ref. \cite{mani-10}, we developed an algorithm to include a class 
of dominant diagrams to all order in $T$, iteratively, 
in the dressed Hamiltonian. Based on this study, we concluded that 
the terms higher than quadratic in $T$ contribute less than $0.1\%$ 
to the properties. So, in the present work, we truncate 
$\tilde {\mathbf D}$ after the second-order in $T$ and include
$\widetilde {\mathbf D} \approx {\mathbf D} + {\mathbf D}T 
+ T^\dagger {\mathbf D } + T^\dagger {\mathbf D}T$ terms in 
the properties calculation.

%%%%%%%%%%%%%%%% figure 1

%%%%%%%%%%%%%%%%%%%%%%%%%%%%%%%%%%%%%%%%%%%%%%%%%%%%%%%%%%%%%%%%%%%%%%%
%%%%                         Results and discussions               %%%%
%%%%%%%%%%%%%%%%%%%%%%%%%%%%%%%%%%%%%%%%%%%%%%%%%%%%%%%%%%%%%%%%%%%%%%%
\section{Results and discussions}
\label{results}

%%%%%%%%%%%%%%%%%%%%%%%%%%%%%%%%%%%%%%%%%%%%%%%%%%%%%%%%%%%%%%%%%%%%%%%%%%%%%%
%%%%     Subsection:Single-particle basis and convergence of properties   %%%%
%%%%%%%%%%%%%%%%%%%%%%%%%%%%%%%%%%%%%%%%%%%%%%%%%%%%%%%%%%%%%%%%%%%%%%%%%%%%%%

\subsection{Single-particle basis and convergence of properties}

An accurate description of single-electron wave functions and corresponding
energies are crucial to obtain the reliable results using FSRCC theory. 
In the present work, we have used the Gaussian-type orbitals (GTOs) \cite{mohanty-91} 
as the single-electron basis for FSRCC calculations.
The GTOs are used as the finite basis sets in which the single-electron wave 
functions are expressed as a linear combination of the Gaussian-type 
functions (GTFs). More precisely, the GTFs of the large component of 
the wavefunction are expressed as
\begin{equation}
  g^L_{\kappa p} (r) = C^L_{\kappa i} r^{n_\kappa} e^{-\alpha_p r^2},
\end{equation}
where $p = 0$, $1$, $2$, $\ldots$, $N$ is the GTO index and $N$ is the
total number of GTFs. The exponent $\alpha_p$ is further expressed as
$\alpha_0 \beta^{p-1}$, where $\alpha_{0}$ and $\beta$ are the two independent
parameters. The parameters $\alpha_{0}$ and $\beta$ are optimized separately
for each orbital symmetry so that the single-electron wavefunctions and
energies match well with the numerical values obtained from the 
GRASP2K \cite{jonsson-13}. The small components of wavefunctions are 
derived from the large components using the kinetic balance 
condition \cite{stanton-84}.

%%%%%%%%%%%%%%%%%%%%%%%%%%%%%%%%%%%%%%%%%%%%%%%%%%%%%%%%%%%%%%%%%
%%%%%% Table 1: Grasp and B-Spline energy compared with Gaussian
%%%%%%%%%%%%%%%%%%%%%%%%%%%%%%%%%%%%%%%%%%%%%%%%%%%%%%%%%%%%%%%%%
\begin{table}
\begin{center}
\caption{The single-particle and SCF energies (in a.u.) from GTO 
	compared with GRASP2K  and B-Spline results. The 
	optimized $\alpha_0$ and $\beta$ parameter for the even 
	tempered basis used in our calculations are also provided. }
\begin{ruledtabular}
\begin{tabular}{cccc}
  Orbitals  &  GTO   & GRASP2K  & B-Spline \\
\hline
$1s_{1/2}$      &  -3257.41150   &  -3257.40298     &  -3257.41571   \\
$2s_{1/2}$  &   -589.39795   &   -589.39605     &   -589.39835   \\
$2p_{1/2}$  &   -565.00395   &   -565.00364     &   -565.00295   \\ 
$2p_{3/2}$  &   -484.41530   &   -484.41512     &   -484.41548   \\      
$3s_{1/2}$  &   -145.26466   &   -145.26400     &   -145.26472   \\ 
$3p_{1/2}$  &   -134.31668   &   -134.31639     &   -134.31638   \\
$3p_{3/2}$  &   -116.10663   &   -116.10639     &   -116.10665   \\
$3d_{3/2}$  &   -98.32473    &    -98.32446     &    -98.32473   \\
$3d_{5/2}$  &   -94.48445    &    -94.48420     &    -94.48445   \\ 
$4s_{1/2}$  &   -35.49304    &    -35.49280     &    -35.49305   \\
$4p_{1/2}$  &   -30.72284    &    -30.72269     &    -30.72276   \\ 	       
$4p_{3/2}$  &   -26.20738    &    -26.20725     &    -26.20737   \\
$4d_{3/2}$  &   -18.43141    &    -18.43130     &    -18.43141   \\ 
$4d_{5/2}$  &   -17.57859    &    -17.57861     &    -17.57859   \\
$4f_{5/2}$  &    -7.44501    &     -7.44494     &     -7.44502   \\
$4f_{7/2}$  &    -7.25330    &     -7.25331     &     -7.25331   \\ 
$5s_{1/2}$  &    -7.63856    &     -7.63855     &     -7.63857   \\      
$5p_{1/2}$  &    -5.94459    &     -5.94458     &     -5.94458   \\ 
$5p_{3/2}$  &    -5.09665    &     -5.09664     &     -5.09666   \\
$5d_{3/2}$  &    -2.62373    &     -2.62373     &     -2.62373   \\ 
$5d_{5/2}$  &    -2.51852    &     -2.51852     &     -2.51852   \\ 
\hline
	$E_{\rm SCF}$ &  -20910.40152  &    -20910.37469     &  -20910.40151 \\
\hline
        &  $\alpha_0$   & $\beta$  &  GTOs        \\
\hline
  $s$    & 0.00450   & 1.805   & 40 \\
  $p$    & 0.00478   & 1.792   & 38  \\
  $d$    & 0.00605   & 1.855   & 34   \\
  $f$    & 0.00355   & 1.845   & 28   \\
 \end{tabular}
\end{ruledtabular}
\label{grasp-en}
\end{center}
\end{table}

In Table \ref{grasp-en}, we have provided the optimized values of $\alpha_{0}$ 
and $\beta$ parameters for Pb$^{2+}$ and have compared the values of single-electron
and self-consistent field (SCF) energies with GRASP2K \cite{jonsson-13} and 
B-spline \cite{zatsarinny-16} results. 
It is to be mentioned that, the single-electron basis used in the 
properties calculations also incorporates the effects of Breit interaction, 
vacuum polarization and the self-energy corrections. As evident from the table,
the single-particle and SCF energies are in excellent agreement with GRASP2K
and B-spline results. The largest difference at the level of SCF and single-particle 
energies are 0.0001\% and 0.0003\%, respectively.

Since GTOs are a mathematically incomplete basis, convergence of the properties 
results with basis size must be checked to get reliable results using FSRCC. 
To show the convergence of properties results, in Table \ref{basis_conv} of 
Appendix, we have listed the values of electric dipole polarizability and E1 and 
M1 transition reduced matrix elements with increasing basis size. To obtain a 
converged basis, we start with a moderate basis size and add orbitals 
systematically to each symmetry until the change in the properties is less 
than or equal to $10^{-3}$ in respective units. For example, as evident from 
the table, the change in E1 amplitude of $\langle^1S_0|| D ||^1P^o_1\rangle$ 
transition is of the order of $10^{-3}$ a.u. when basis is augmented from 
158 ($24s21p18d13f8g7h$) to 169 ($25s22p19d14f9g8h$) orbitals. So, to minimize 
the computation time, we consider the basis set with 169 orbitals as 
optimal, and use it for further FSRCC calculations where the corrections
from the Breit interaction, vacuum polarization and the self-energy 
are incorporated.

%%%%%%%%%%%%%%%% figure 1
\begin{figure}
 \includegraphics[scale=0.28, angle=-90]{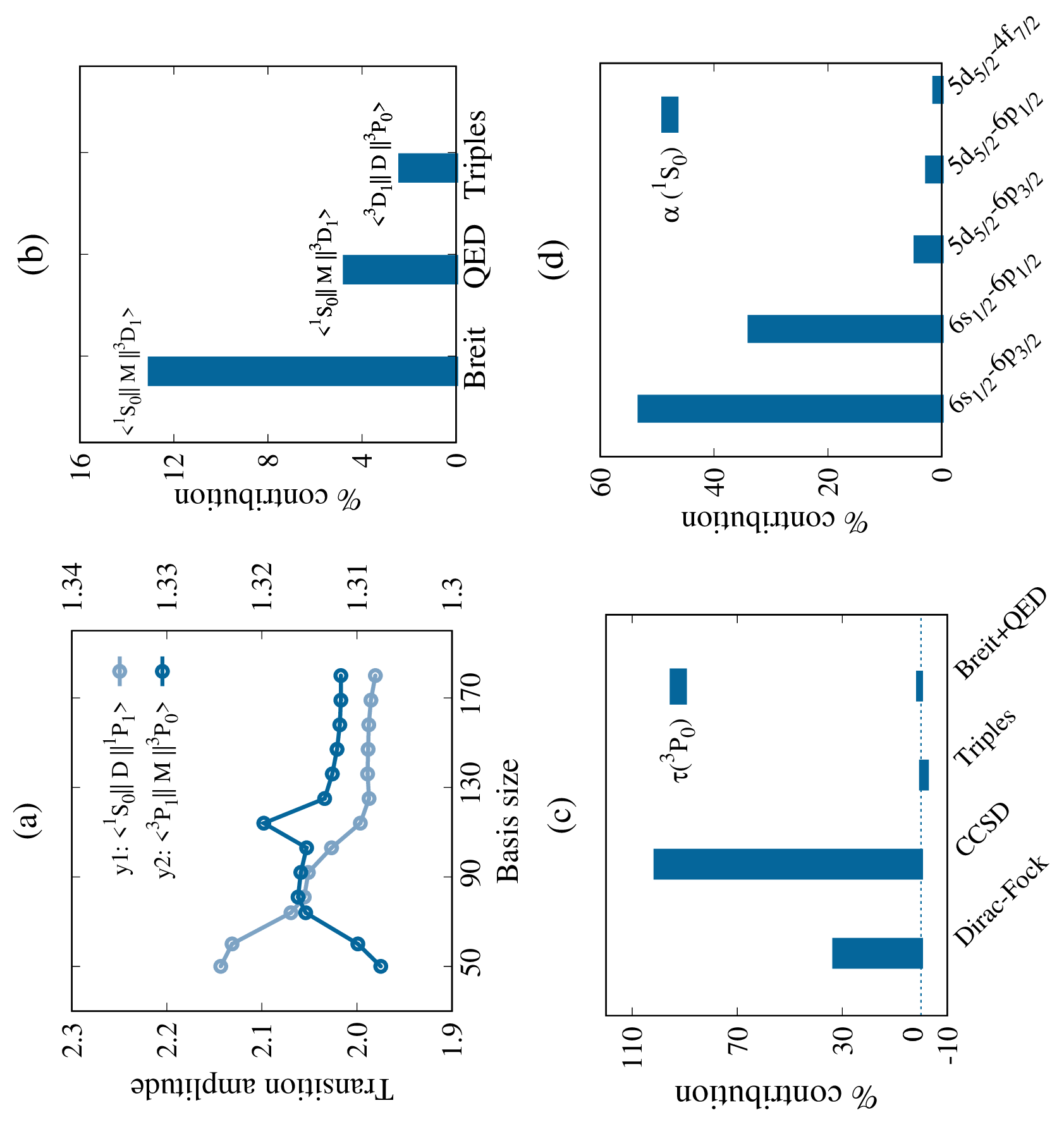}
	\caption{Convergence of transition amplitudes with basis size, panel (a). 
	Dominant percentage contributions from Breit, perturbative triples 
	and QED corrections to the reduced matrix elements, panel (b), and 
	the lifetime of the clock state, panel (c).  Dominant contributions 
	from core orbitals to the dipole polarizability, panel (d).}
 \label{fig_tbq}
\end{figure}

%%%%%%%%%%%%%%%%%%%%%%%%%%%%%%%%%%%%%%%%%%%%%%%%%%%%%%%%%%%%%%%%%%%%%%%%%%%%%%
%%%%                      Subsection:Excitation Energy                    %%%%
%%%%%%%%%%%%%%%%%%%%%%%%%%%%%%%%%%%%%%%%%%%%%%%%%%%%%%%%%%%%%%%%%%%%%%%%%%%%%%
\subsection{Excitation Energy}

The eigen energies obtained from the solution of many-electron
Schrodinger equation, Eq. (\ref{hdc_2v}), using FSRCC are used to calculate 
the excitation energies. The excitation energy of a general state 
$nln'l'{\;^{(2S+1)}}L_J$ is defined as 
\begin{equation}
        \Delta E_{nln^{'}l^{'}{\;^{(2S+1)}}L_J} =
        E_{nln^{'}l^{'}{\;^{(2S+1)}}L_J} - E_{ns^2{\;^1}S_0},
\end{equation}
where $E_{ns^2{\;^1}S_0}$ and $E_{nln^{'}l^{'}{\;^{(2S+1)}}L_J}$
are the exact energies of the ground and excited states, respectively.
In Table \ref{tab-ee}, we have listed the excitation energies from our 
calculations along with other theoretical and experimental data for comparison.
To account for {\em valence-valence} correlations more accurately, we have 
also included $6p^{2}$, $6s6d$ and $6s7s$ configurations in 
the model space. For a quantitative assessment of electron correlations, 
we have listed the contributions from Breit and QED corrections separately.

%%%% Table: Basis-173($6s^{2}$ + $6s6p$ + $6p^{2}$ + $6s6sd$ + $6s7s$) 
\begin{table*}
  \caption{The energy of ground state $^1S_0$ (cm$^{-1}$) and excitation 
	energies of some low lying excited states of Pb$^{2+}$. For quantitative 
	analysis of electron correlations, contributions from
	Breit and QED corrections are given separately.}
\begin{ruledtabular}
\begin{tabular}{crrrrrrrr}
\multicolumn{1}{c}{\textrm{States}}  &
\multicolumn{1}{c}{\text{DC-CCSD}}&
\multicolumn{1}{c}{\text{Breit}}&
\multicolumn{1}{c}{\text{Self-energy}}&
\multicolumn{1}{c}{\text{Vac-pol}} &
\multicolumn{1}{c}{\text{Total}}&
\multicolumn{1}{c}{\text{Other cal.}} & 
\multicolumn{1}{c}{\text{NIST\cite{NIST-ASD}}} &     
\multicolumn{1}{c}{\text{\% Error}}  \\                                  
\hline     
% \multicolumn{10}{c}{Excitation energies (in $cm^{-1}$)} \\ \\
$6s^{2}$\ $ ^1S_0$     &  599355.44  &   12.00 & -0.67 & 188.97 &  599556  & 600984\cite{safronova-12}                          & 598942 & 0.1   \\ 

$6s6p$\ $ ^3P^{o}_0$   &   59624.80  &  121.57 & -0.58 &  81.41 &   59827  &  61283\cite{safronova-12}, 60653\cite{curtis-01}   &  60397 & 0.94  \\
$6s6p$\ $ ^3P^{o}_1$   &   64146.99  &  116.92 & -0.53 &  82.84 &   64346  &   65089\cite{safronova-12}, 65683\cite{migdalek-85}&  64391 & 0.06  \\
                       &             &         &       &        &          &    60387\cite{migdalek-88},58905\cite{chou-92}     &        &       \\
                       &             &         &       &        &          &    64609\cite{curtis-01}                           &        &        \\
$6s6p$\ $ ^3P^{o}_2$   &   79478.69  &  83.67  & -0.33 &  90.27 &   79652  &   80029\cite{safronova-12}, 79024\cite{curtis-01}  &  78985 & -0.8   \\      
$6s6p$\ $ ^1P^{o}_1$   &   95045.89  &  83.07  & -0.63 &  84.30 &   95213  &   95847\cite{safronova-12}, 97970\cite{migdalek-85}&  95340 & 0.13   \\ 
                       &             &         &       &        &          &    91983\cite{migdalek-88}, 95537\cite{chou-92}    &        &        \\
                       &             &         &       &        &          &    95535\cite{curtis-01}                           &        &        \\
$6p^{2}$\ $ ^3P_0$     &  142922.33  & 236.01  & -1.41 & 169.97 &  143327  &  143571\cite{safronova-12}                         & 142551 & -0.54  \\
$6s7s$\ $ ^3S_1$       &  149898.62  &   4.90  & -0.32 &  64.10 &  149967  &  151183\cite{safronova-12}                         & 150084 &  0.07  \\
$6s6d$\ $ ^1D_2$       &  152651.06  & 108.27  & -0.70 & 135.00 &  152894  &  153614\cite{safronova-12}                         & 151885 & -0.6   \\
$6s7s$\ $ ^1S_0$       &  153901.92  &   3.47  & -0.28 &  62.88 &  153968  &  155054\cite{safronova-12}                         & 153783 & -0.12  \\
$6p^{2}$\ $ ^3P_1$     &  155401.22  & 189.27  & -1.16 & 170.87 &  155760  &  156610\cite{safronova-12}                         & 155431 & -0.2   \\
$6s6d$\ $ ^3D_1$       &  157523.12  &  17.37  & -0.55 &  92.43 &  157632  &  158439\cite{safronova-12}                         & 157444 & -0.1   \\
$6s6d$\ $ ^3D_2$       &  157902.47  &   5.93  & -0.48 &  87.81 &  157996  &  159134\cite{safronova-12}                         & 157925 & -0.04  \\
$6s6d$\ $ ^3D_3$       &  159147.60  &   0.04  & -0.38 &  86.03 &  159233  &  160530\cite{safronova-12}                         & 158957 & -0.17  \\
$6p^{2}$\ $ ^3P_2$     &  164987.49  & 117.07  & -0.99 & 143.34 &  165247  &  165898\cite{safronova-12}                         & 164818 & -0.26  \\
$6p^{2}$\ $ ^1D_2$     &  179412.09  & 139.08  & -1.17 & 166.70 &  179717  &  179646\cite{safronova-12}                         & 178432 & -0.7   \\
$6p^{2}$\ $ ^1S_0$     &  189714.06  & 161.46  & -1.17 & 179.67 &  190054  &  190061\cite{safronova-12}                         & 188615 & -0.7   \\    
\end{tabular}
\end{ruledtabular}
\begin{flushleft}
 $^{\rm a}$  Ref. \cite{safronova-12} - CI + all-order , 
 $^{\rm b}$ Ref. \cite{curtis-01} - MCDHF, 
 $^{\rm c}$ Ref. \cite{migdalek-85} - CIRHF + CP, 
 $^{\rm d}$ Ref. \cite{migdalek-88} - CIRHF + CP, 
 $^{\rm e}$ Ref. \cite{chou-92} - MCRRPA
\end{flushleft}
\label{tab-ee}
\end{table*}

As evident from the table, our computed energies are in excellent agreement 
with the experimental results. The largest relative error in our calculation 
is $\approx$ 0.9\%, which corresponds to the ${^3}P^{o}_0$ state. However, for 
other states, specially for those which contribute to the lifetime of the 
clock state, the errors are much smaller. The states 
${^3}P^{o}_1$, ${^1}P^{o}_1$, ${^3}S^{}_1$ and ${^3}D^{}_1$, which couple 
either via E1 or M1 operator in the clock transition, have the relative errors 
of 0.06\%, 0.13\%, 0.07\% and -0.10\%, respectively. This is crucial, as these 
energies contribute to the lifetime of the clock state. Among all the previous 
theoretical results listed in the table, 
Ref. \cite{safronova-12} is close to ours in terms of the many-body methods 
used, however, with an important difference. Ref. \cite{safronova-12} uses 
a {\em linearized} CCSD method, whereas the present work employs a 
{\em nonlinear} CCSD, which accounts for electron correlation effects 
more accurately in the calculation. The relative errors in the reported 
excitation energies for ${^3}P^{o}_1$, ${^1}P^{o}_1$, ${^3}S^{}_1$ 
and ${^3}D^{}_1$ states in Ref. \cite{safronova-12} are  1.08\%, 0.53\%, 
0.73\% and 0.63\%, respectively. Remaining results are mostly based 
on the multi-configuration Hartree-Fock and its variations and, 
in general, not consistent in terms of treating electron correlations. 

Examining the contributions from high energy configurations, we observed
an improvement in the excitation energies of ${^3}P^{o}_1$ and ${^1}P^{o}_1$ 
states due to accounting of {\em valence-valence} correlation more accurately.
We find that the relative error has reduced from 0.7\% (0.6\%) to 0.4\% (0.3\%)
for ${^3}P^{o}_1$(${^1}P^{o}_1$) state. Among the contributions from Breit, 
vacuum polarization and self-energy corrections, the former two are observed 
to contribute more. The largest cumulative contribution of 
about $0.3$\% from Breit and vacuum polarization is observed in the case of 
$^3P^o_0$. Self-energy contributions are of opposite phase and negligibly small.

%%%%%%%%%%%%%%%%%%%%%%%%%%%%%%%%%%%%%%%%%%%%%%%%%%%%%%%%%%%%%%%%%%%%%%%%%%%%%%
%%%%             Subsection:E1 Reduced Matrix Elements                    %%%%
%%%%%%%%%%%%%%%%%%%%%%%%%%%%%%%%%%%%%%%%%%%%%%%%%%%%%%%%%%%%%%%%%%%%%%%%%%%%%%
\subsection{E1 Reduced Matrix Elements}

In Table \ref{tab-e1m1}, we have listed the values of E1 reduced 
matrix elements from our calculations for all dominant transitions 
which contribute to the lifetime of the clock state. Since there are 
more data on oscillator strengths in the literature, we have converted E1 
reduced matrix elements to oscillator strength, and tabulated in the table for 
comparison with experiments and other theoretical results. 
The contributions from Breit+QED and triples are provided separately 
in the table. As evident from the table and as to be expected, DC-CCSD
is the dominant contribution to all the matrix elements. The contributions 
from Breit, QED and perturbative triples are important to obtain accurate results. 
A quantitative analysis is presented later in the section.

%%%%%%%%%%%%%%%%%%%%%%
\begin{table*}
	\caption{The E1 and M1 reduced matrix elements (a.u.) and oscillator
	strengths for some allowed transitions in Pb$^{2+}$. For comparison, data
	from experiments and other theory calculations are also provided.}
\begin{ruledtabular}
\begin{tabular}{ccccccc}
\multicolumn{1}{c}{\textrm{States}}  &
\multicolumn{1}{c}{\text{DC-CCSD}}&
\multicolumn{1}{c}{\text{Breit + QED}}&
\multicolumn{1}{c}{\text{P-triples}}&
\multicolumn{1}{c}{\text{Total}}&
\multicolumn{1}{c}{\text{Other cal.}}&
\multicolumn{1}{c}{\text{Expt.}} \\
\hline
%%%%%%%%%%%%%%%%%%%%%%%%%%%%%%%%%%
\multicolumn{6}{c}{E1 Reduced Matrix Elements} \\
$\langle^1S_0|| D ||^3P^o_1\rangle$ &  0.5319 &   -0.0007 &  -0.0024  & 0.5288 & 0.706\cite{beloy-21}, 0.644\cite{safronova-12} &  \\
$\langle^1S_0|| D ||^1P^o_1\rangle$ &  1.9854 &   0.0003 &  -0.0327  & 1.9530 & 2.350\cite{beloy-21}, 2.384\cite{safronova-12}  & \\
$\langle^3S_1|| D ||^3P^o_0\rangle$ &  0.5362 &   0.0120 &  -0.0123  & 0.5359 &  0.963\cite{safronova-12} & \\
$\langle^3D_1|| D ||^3P^o_0\rangle$ & -1.4796 &   0.0171 &  -0.0355  & -1.4980 & -1.516\cite{safronova-12}  & \\  \\

%%%%%%%%%%%%%%%%%%%%%%%%%%%%%%%%%%%
	\multicolumn{6}{c}{Oscillator Strengths} \\
$\langle^1S_0|| D ||^3P^o_1\rangle$ & 5.5071[-2] & 0.0026[-2] & -0.4950[-2] &  5.4602[-2]
	& 8.11[-2]\cite{safronova-12}, 7.40[-2]\cite{alonso-09}, 5.44[-2]\cite{migdalek-85},
	&  (7.3 $\pm$ 0.5)[-2]\cite{pinnington-88}, \cite{ansbacher-88} \\
	&  &&   &   &  7.55[-2]\cite{glowacki-03}, 6.15[-2]\cite{chou-92}, 6.15[-2]\cite{chou-97} &      \\
	&  &&   &   &  5.52[-2]\cite{migdalek-88}  & \\

$\langle^1S_0|| D ||^1P^o_1\rangle$ &  1.1369 & 0.0023  &  -0.0371  & 1.1021
	&  1.65\cite{safronova-12}, 1.24\cite{alonso-09}, 1.64\cite{migdalek-85}, 1.51\cite{glowacki-03},
	&  (1.01 $\pm$ 0.20)\cite{andersen-72}  \\
	&   &   &   &      & 2.45\cite{chou-92}, 1.42\cite{chou-97}, 1.43\cite{migdalek-88} &  \\

$\langle^3S_1|| D ||^3P^o_0\rangle$ & 0.0788 &   0.0034      & -0.0036     & 0.0786  & 0.229\cite{alonso-09}   & \\
$\langle^3D_1|| D ||^3P^o_0\rangle$ & 0.6504 & -0.0156  &  0.0315     & 0.6663  & 0.93\cite{alonso-09}  &  \\ \\

%%%%%%%%%%%%%%%%%%%%%%%%%%%%%%%%%%
\multicolumn{6}{c}{M1 Reduced Matrix Elements} \\
$\langle^3P^o_1|| M1 ||^3P^o_0\rangle$ &  -1.3117  &    -0.0005 & 0.0006  & -1.3116
	&  -0.674\cite{beloy-21}  & \\
$\langle^1P^o_1|| M1 ||^3P^o_0\rangle$ & 0.4972  &   0.0002 & 0.0001  & 0.4975
	&  0.205\cite{beloy-21} & \\
$\langle^1S_0|| M1 ||^3S_1\rangle$ & 0.0044  &   -0.0003 &  0   & 0.0041
	&    & \\
 $\langle^1S_0|| M1 ||^3D_1\rangle$ & -0.0143  &  -0.003 & 0.0004  & -0.0169
	&    & \\
\end{tabular}
\end{ruledtabular}
\begin{flushleft}
  $^{\rm a}$  Ref. \cite{beloy-21} - CI + MBPT ,
  $^{\rm b}$ Ref. \cite{alonso-09} - IC + RHF +CP,
  $^{\rm c}$ Ref. \cite{glowacki-03} - CIDF - MP,
  $^{\rm d}$ Ref. \cite{chou-97} - MCRRPA,
  $^{\rm e}$ Ref. \cite{safronova-12} - CI + all-order,
  $^{\rm f}$ Ref. \cite{colon-00} - IC + RHF,
  $^{\rm g}$ Ref. \cite{migdalek-88} - CIRHF + CP,
  $^{\rm h}$ Ref. \cite{chou-92} - MCRRPA,
  $^{\rm i}$ Ref. \cite{migdalek-85} - CIRHF + CP,
%  $^{\rm i}$ Ref. \cite{chou-97} - MCRRPA
\end{flushleft}
\label{tab-e1m1}
\end{table*}

From the literature we could find two previous
works, Refs. \cite{beloy-21}-CI+MBPT and \cite{safronova-12}-CI+all-order, 
for comparison of the E1 reduced matrix elements. 
The values of our E1 reduced matrix elements are slightly smaller than 
Refs. \cite{beloy-21} and \cite{safronova-12} for all the listed transitions. 
The reason for this could be attributed to the different treatment of electron 
correlations in these methods. In Ref. \cite{beloy-21}, MBPT is used to 
treat {\em core-core} and {\em core-valence} correlations, whereas 
{\em valence-valence} correlation is incorporated with CI. In 
Ref. \cite{safronova-12}, however, the {\em core-core} and {\em core-valence} 
correlations are accounted using a {\em linearized} CCSD method.
The present work, however, employs a {\em nonlinear} CCSD theory to account 
for the {\em core-core} and {\em core-valence} correlations and, therefore, 
is more accurate. The {\em valence-valence} correlation is however treated 
in the same way as in Refs. \cite{beloy-21, safronova-12}. The other two 
important inclusions in the present work are, the use of energetically higher 
configurations ($6p^2$, $6s7s$ and $6s6d$) in the model space and the 
corrections from the Breit, QED and perturbative triples.

For the oscillator strength, there are several results in the literature from
the previous studies for the $\langle^1S_0||D||^3P^o_1\rangle$ and 
$\langle^1S_0||D||^1P^o_1\rangle$ transitions for comparison. 
As evident from the table, there is, however, a large variation in the reported 
values. For example, for the $\langle^1S_0||D||^3P^o_1\rangle$ transition, the 
lowest result $5.44\times10^{-2}$ from Ref. \cite{migdalek-85} differs by 
$\approx$ 33\% from the highest result $8.11\times10^{-2}$ reported in the 
Ref \cite{safronova-12}. A similar trend is also observed for the
$\langle^1S_0||D||^1P^o_1\rangle$ transition. The lowest value, 
1.24 \cite{alonso-09}, is close to half the highest value, 
2.45 \cite{chou-92}. The reason for the large variation could be attributed 
to the different many-body methods employed in these calculations. It is to 
be noted that, none of the previous calculations use FSRCC theory, like in the 
present work. Except Ref. \cite{safronova-12}, which uses CI + all-order, 
the other calculations are mostly based on the MCDF and its variations. The 
large difference among the results clearly indicates the inherent dependence
of the results on the choice of configurations in the MCDF method to 
incorporate electron correlation effects. For the 
$\langle^1S_0||D||^3P^o_1\rangle$ transition, our result, $5.46\times10^{-2}$, 
lies within the range of the previous results. Whereas, for 
$\langle^1S_0||D||^1P^o_1\rangle$ transition, our result, 1.10, is 
lowest among all the results listed in the table.

From experiments, to the best of our knowledge, there is one result each 
for oscillator strength for $\langle^1S_0||D||^3P^o_1\rangle$ \cite{pinnington-88} 
and $\langle^1S_0||D||^1P^o_1\rangle$ \cite{andersen-72} transitions. 
Both of these experiments use the beam-foil technique to study atomic 
spectra. For  $\langle^1S_0||D||^3P^o_1\rangle$ transition, our calculated
result, $5.46\times 10^{-2}$, has the same order of magnitude as the 
experimental result, $(7.3 \pm 0.5)\times 10^{-2}$, but about 25\% smaller.
Among the previous theory calculations, the MCDF calculations Refs. \cite{alonso-09} 
and \cite{glowacki-03} are closer to the experiment.
For $\langle^1S_0||D||^1P^o_1\rangle$ transition, however, among all the
theory results listed in table, our result, 1.10, has the best match 
with the experimental result, $1.01 \pm 0.20$ \cite{andersen-72}.

%%%%%%%%%%%%%%%%%%%
\begin{table}
    \caption{Termwise contributions to E1 and M1 reduced matrix
	elements (a.u.) from different terms in FSRCC theory. The
	operator $\hat O$ represents the electric or magnetic dipole
	operator.}
    \label{pol_tw}
    \begin{center}
    \begin{ruledtabular}
    \begin{tabular}{lrrrr}
%        \hline
        Terms + h. c. & $\langle^1S_0|| D ||^1P^o_1\rangle$ & $\langle^3P^o_1|| M1 ||^3P^o_0\rangle$ \\
        \hline
        DF                   & $2.1575$ & $-1.3586$   \\
        1v diagrams          & $-0.4014$  & $-0.0122$     \\
        $\hat O R_{2}$             & $0.1064$ & $0.0199$      \\
        ${R_{2}}\hat O {R_{2}}$    & $0.0735$ & $0.0423$  \\
        ${S_{1}}\hat O {R_{2}}+{S_{2}}\hat O {R_{2}}+{S_1^{2}}\hat O {R_{2}}$& $0.0101$  & $-0.0081 $ \\
        $\hat O S_{2} $   & $0.0065$ & $0.0003$  \\
        ${S_{2}}\hat O {S_{2}}+{S_{2}} \hat O {S_{1}}+{S_{2}}\hat O {S_1^{2}}$& $0.0320$  & $0.0046 $ \\
        ${T_{1}}\hat O {R_{2}}$    & $-0.0001$ & $0$  \\
        ${S_{2}}\hat O {T_{2}}+{S_{2}}\hat O{T_{1}}$& $0.0007$  & $0 $ \\
        ${T_{2}} \hat O {T_{2}}+{T_{1}} \hat O {T_{2}}$& $0.0003$  & $0 $ \\
        Total    & $1.9855$  & $-1.3118$ \\
    \end{tabular}
    \end{ruledtabular}
    \end{center}
    \label{term-e1m1}
\end{table}
%%%%%%%%%%%%%%%

%%%%%%%%%%%%%%%%%%%%%%%%%%%%%%%%%%%%%%%%%%%%%%%%%%%%%%%%%%%%%%%%%%%%%%%%%%%%%%
%%%%             Subsection:M1 Reduced Matrix Elements                    %%%%
%%%%%%%%%%%%%%%%%%%%%%%%%%%%%%%%%%%%%%%%%%%%%%%%%%%%%%%%%%%%%%%%%%%%%%%%%%%%%%
\subsection{M1 Reduced Matrix Elements}

 For the clock transition, the theoretical estimate of the M1 matrix elements is 
the other important component to calculate the lifetime of the clock state. 
So, we next compute the M1 reduced matrix elements of the transitions 
which contribute dominantly to $\tau$. These are listed in the 
Table \ref{tab-e1m1}. As to be expected, like the case of E1 matrix elements, 
the most dominant contribution is from DC-CCSD for all the transitions.  
The cumulative contribution from Breit, QED and perturbative triples are small 
but important to get reliable transition properties results.

Unlike the E1 matrix elements, only few results of M1 are available in the
literature for comparison. To the best of our knowledge, there is only one 
theoretical result calculated using CI+MBPT \cite{beloy-21}, which reports 
the values of M1 reduced matrix elements for the
$\langle^3P^o_1|| M1 ||^3P^o_0\rangle$ and 
$\langle^1P^o_1|| M1 ||^3P^o_0\rangle$ transitions. Interestingly, unlike the 
E1 reduced matrix elements where the two works are comparable, our results 
for M1 reduced matrix elements differ by a factor of two or more from 
Ref. \cite{beloy-21}. This leads to a difference of $\approx$ 6.7\% 
between the lifetimes calculated using the two data. Our calculated value of 
$\tau$ with $6s^2 + 6s6p$ configuration is $9.6 \times 10^6$ s, whereas the 
value reported in Ref. \cite{beloy-21} is $9.0 \times 10^6$ s.
As this work reports our first implementation and computation of M1 matrix 
element for two-valence system using FSRCC, it is essential to cross check and 
validate our results with previous works. For this we compute and compare the
results of other atoms since there are no previous theoretical or experimental 
results for Pb$^{2+}$ other than Ref. \cite{beloy-21}. In particular, we consider 
the  M1 transition rate for the $\langle^3P^o_2|| M1 ||^3P^o_1\rangle$ transition in neutral 
Yb. This was studied in Ref. \cite{dzuba-18}, using a combined method of 
configuration interaction (CI) and perturbation theory (PT), and reported 
a value as $6.7\times10^{-2}$ s$^{-1}$. And from our implementation we obtain
$5.3\times10^{-2}$ s$^{-1}$. Reason for this small difference could be attributed 
to the better consideration of electron 
correlations in FSRCC theory. In another work Ref. \cite{derevianko-01} 
the transition rate for $\langle^3P^o_2|| M1 ||^3P^o_1\rangle$ 
of Sr was computed using CI+RPA. It reported the value as $8.26\times 10^{-4}$, 
this matches very well with our result of $8.88\times 10^{-4}$. The difference
is only 7\% which could again be due to better accounting of electron 
correlations in FSRCC. In yet another seminal work, Ref. \cite{safronova-99} 
carried out a second-order MBPT calculation of M1 reduced matrix elements for 
$\langle^3P^o_0|| M1 ||^3P^o_1\rangle$ and 
$\langle^3P^o_0|| M1 ||^1P^o_1\rangle$ transitions in Fe$^{22+}$. The reported 
values, 1.40 and 0.22, respectively are in good agreement with our computed 
values, 1.37 and 0.28. Thus, from all the comparison with the previous 
theoretical results for different systems, we can infer that our 
implementation of M1 matrix element computation with FSRCC gives reliable
results.

%%%%%%%%%%%%%%%%%%%%%%%%%%%%%%%%%%%%%%%%%%%%%%%%%%%%%%%%%%%%%%%%%%%%%%%%%%%%%%
%%%%                Subsection:Lifetime of Clock State                    %%%%
%%%%%%%%%%%%%%%%%%%%%%%%%%%%%%%%%%%%%%%%%%%%%%%%%%%%%%%%%%%%%%%%%%%%%%%%%%%%%%
\subsection{Lifetime of Clock State}

The lifetime of the clock state can now be theoretically estimated using
the results discussed above. Using the  E1 and M1 reduced matrix elements listed in 
Table \ref{tab-e1m1} and excitation energies from Table \ref{tab-ee} in 
Eq. (\ref{e1m1}), we obtain the E1M1 decay rate ($\Gamma$) for the 
$6s^2{\;^1}S_0 \rightarrow 6s6p{\;^3}P^o_0$ clock transition and its inverse  is
$\tau$. The $\tau$ obtained from our calculations is given in the 
Table \ref{tab-tau}. To assess the effect of {\em valence-valence} correlation,
we have separated the contributions from $6s6p$, $6s7s$ and $6s6d$ 
configurations. Our computed lifetime, $9.76 \times 10^{6}$ s,
is $\approx 8.5$\% larger than the only other theoretical result \cite{beloy-21}.
The reason for this could be attributed to the more accurate treatment of 
electron correlations in our calculations. It should be noted that, our 
calculations also incorporate the contributions from E1 and M1 matrix elements 
from higher energy states $6s6d{\;^3}D_1$ and $6s7s{\;^3}S_1$. 
This has a significant cumulative contribution of $\approx$ 3.3\% to the total 
lifetime. The other key difference from Ref. \cite{beloy-21} is, the inclusion 
of the corrections from Breit, QED and perturbative triples in our calculations. 
The contribution from the perturbative triples is $\approx -2.3$\% to the lifetime. 
This is consistent with the trend observed in our previous work on Al$^+$
atomic clock \cite{ravi-21a}. The contributions from the Breit and QED corrections 
are also significant, they jointly contribute $\approx 1.1$\% of 
the total $\tau$. As discernible from the Fig. \ref{fig_tbq}(c), the DF 
alone contributes $\approx 33$\% of the total value. The most significant 
contribution arises from the electron correlations associated with the residual 
Coulomb interaction through FSRCC within the CCSD framework. The combined 
contribution from DF and CCSD is about 101\% of the total lifetime. 

%%%%%%%%%%%%%%%%%%%%
\begin{table}
    \caption{The lifetime of the clock state $^3P^{o}_0$. The contributions 
	from the Breit, QED and perturbative triples corrections are provided 
	separately.}
    \label{decay_rate}
    \begin{center}
    \begin{ruledtabular}
    \begin{tabular}{lc}
%        \hline
        Configurations/Methods &  $\tau$ ($\times 10^{6}$ s) \\
        \hline
	   6s$^2$ + 6s6p            & $9.595 $  \\
	   6s7s                     & $0.029 $  \\
	   6s6d                     & $0.248 $  \\
	   Total CCSD                & $9.872 $  \\
	   CCSD(T)                  & $9.654 $  \\
	   CCSD(T)+Breit+QED        & $9.761 $  \\
	   Recommended              & $(9.76 \pm 0.47) $  \\
      Others                        & $9.0$\cite{beloy-21}  \\
    \end{tabular}
    \end{ruledtabular}
    \end{center}
    \label{tab-tau}
\end{table}
%%%%%%%%%%%%%%

%%%%%%%%%%%%%%%%%%%%
%\begin{table}
%    \caption{E1M1 decay rate and lifetime of the clock state $^3P^{o}_0$. Tabulated results
%	also incorporate the contributions from the Breit, QED and perturbative triples
%	corrections.}
%    \label{decay_rate}
%    \begin{center}
%    \begin{ruledtabular}
%    \begin{tabular}{lcc}
%%        \hline
%        Configurations & $\Gamma$ $\times 10^{-8} $ (sec$^{-1}$) & $\tau$ $\times 10^{6}$(sec) \\
%        \hline
%	   6s$^2$ + 6s6p               & $10.422 $ & $9.595 $  \\
%	   6s7s                        & $-0.031 $ & $0.029 $  \\
%    % 6s$^2$ + 6s6p + \bf{6s6d}   & $10.179 $ & $9.824 $  \\
%	     6s6d                        & $-0.263 $ & $0.248 $  \\
%	     Total CCSD               & $10.128 $ & $9.872 $  \\
%	     CCSD(T)                  & $10.359 $ & $9.654 $  \\
%	     CCSD(T)+Breit+QED        & $10.245 $ & $9.761 $  \\
%	     Recommended                 & $10.245 $ & $(9.761 \pm 0.47) $  \\
%      Others                      &           & $9.0$\cite{beloy-21}  \\
%    \end{tabular}
%%    \end{ruledtabular}
%    \end{center}
%    \label{tab-tau}
%\end{table}
%%%%%%%%%%%%%%%

%%%%%%%%%%%%%%%%%%%%%%%%%%%%%%%%%%%%%%%%%%%%%%%%%%%%%%%%%%%%%%%%%%%%%%%%%%%%%%
%%%%                  Subsection:Dipole Polarizability                    %%%%
%%%%%%%%%%%%%%%%%%%%%%%%%%%%%%%%%%%%%%%%%%%%%%%%%%%%%%%%%%%%%%%%%%%%%%%%%%%%%%
\subsection{Dipole Polarizability}

The electric dipole polarizability, $\alpha$, of an atom or ion is a measure of 
the response to an external electric field. It is related to properties which
serve as signatures of several fundamental properties 
\cite{khriplovich-91, griffith-09,karshenboim-10}. In the 
present work, $\alpha$ is required in calculating the BBR shift of the 
clock transition frequency. In the Table \ref{tab-alpha}, we present our 
theoretical result on $\alpha$ for the ground state of Pb$^{2+}$ and compare
it with results available in the literature. To calculate $\alpha$, 
we have used the PRCC theory developed and presented in our previous 
works \cite{ravi-20, ravi-21b}. The Table \ref{tab-alpha} also list the 
contributions from various correlation terms subsumed in the PRCC theory. The 
term {\em estimated} identifies the contribution from the orbitals from 
$i$, $j$ and $k$-symmetries. As to be expected, the dominant contribution is 
from the DF term. It contributes $\approx 116$\% of the total value. The 
PRCC value is $\approx 13$\% lower than the DF value. The reason for this is 
the cancellation due to opposite contributions from electron correlation.

%%%%%%%%%%%%%%%%%%%% Table: Polarizability  %%%
\begin{table}
    \caption{The value of $\alpha$ (a.u.) for ground state, $6s^2\;^1S_{0}$, 
	of Pb$^{2+}$ from PRCC calculation. The available data from experiment 
	and other theory calculations are also provided for comparison.}
  \begin{ruledtabular}
  \begin{tabular}{clcr}
      State & \multicolumn{2}{c}{\textrm{Present work}} \\
            \cline{2-3}
           & {\textrm{Method}}& $\alpha$ &  \\
            \hline
$6s^2\;^1S_{0}$  & DF                & $16.246$  \\
                 & PRCC              & $14.173$  \\
                 & PRCC(T)           & $14.166$  \\
	         & PRCC(T)+Breit     & $14.173$   \\
                 & PRCC(T)+Breit+QED & $14.064$  \\
                 & Estimated         & $14.016$  \\
                 & Recommended       & $14.02 \pm 0.21$  \\
                 & Other cal. & $13.3 \pm 0.4^{\rm a}$    \\
		 & Expt.      & $13.62\pm 0.08^{\rm b}$ \\
        \end{tabular}
        \end{ruledtabular}
\begin{tabbing}
  $^{\rm a}$Ref.\cite{safronova-12}-CI + all-order,  
  $^{\rm b}$Ref.\cite{hanni-10}-Expt. 
%  $^{\rm c}$Ref.\cite{reshetnikov-07}-Expt. 
\end{tabbing}
\label{tab-alpha}
\end{table}

%%%%%%%%%%%%%%%%%%%%%
\begin{table}
    \caption{Termwise contributions to $\alpha$ (a.u.) from different 
	     terms in PRCC theory.}
    \begin{center}
    \begin{ruledtabular}
    \begin{tabular}{cc}
%        \hline
        Terms + h. c. & $\alpha$  \\
        \hline                                                 
	    ${{\mathbf T}_{1}^{(1)\dagger}{\mathbf D}} $                  & $16.6849$   \\
        ${\mathbf T_{1}}{^{(1)\dagger}}{\mathbf D}T_{2}^{(0)} $   & $-1.0835$   \\
        ${\mathbf T_{1}}{^{(1)\dagger}}{\mathbf D}T_{1}^{(0)} $   & $-0.2746$    \\
        ${\mathbf T_{2}}{^{(1)\dagger}}{\mathbf D}T_{1}^{(0)} $   & $-0.0052$     \\
        ${\mathbf T_{2}}{^{(1)\dagger}}{\mathbf D}T_{2}^{(0)} $   & $0.4145$      \\
        Normalization                             & $-1.5633$        \\
        Total                                     & $14.1733$         \\
    \end{tabular}
    \end{ruledtabular}
    \end{center}
    \label{pol_tw}
\end{table}

From the literature, we could find one result each from the experimental and
theoretical studies. On comparing the results, our recommended value, 14.02, is 
in good agreement with the experimental value, 13.62$\pm$0.08, reported 
in Ref. \cite{hanni-10}. The difference from the experimental result is 
$\approx 3$\%. In theoretical work of Safronova and collaborators 
\cite{safronova-12}, the reported value of 13.3 is obtained using the
CI+all-order method.  Our recommended result is $\approx 6$\% larger than Ref. 
\cite{safronova-12}. As mentioned earlier, reason for this 
difference could be attributed to the more accurate treatment of electron 
correlations in the present calculation. The other important advantage of 
the present calculation is that, it does not employ the 
sum-over-state approach \cite{safronova-99a, derevianko-99} to incorporate 
the effects of perturbation. The summation over all the possible intermediate 
states is accounted through the perturbed cluster 
operators \cite{ravi-20, ravi-21a}. In addition, the present work also 
incorporates the effects of Breit, QED and perturbative triples corrections 
in the calculation of $\alpha$.

%%%%%%%%%%%%%%%%%%%%%%%%%%%%%%%%%%%%%%%%%%%%%%%%%%%%%%%%%%%%%%%%%%%%%%%%%%%%%%
%%%%                Subsection:Electron Correlations in FSRCC...          %%%%
%%%%%%%%%%%%%%%%%%%%%%%%%%%%%%%%%%%%%%%%%%%%%%%%%%%%%%%%%%%%%%%%%%%%%%%%%%%%%%
\subsection{Electron Correlations in FSRCC and PRCC Theories 
            and Corrections from Breit, QED and Perturbative Triples}

To get insights on the correlation effects, we now analyze and present the trend 
of contributions from various correlation terms in FSRCC and PRCC theories as
well as the contributions from the Breit and QED corrections. As mentioned 
earlier, the FSRCC method is used to calculate the lifetime of the
metastable clock state, whereas PRCC theory is employed to calculate 
$\alpha$ for the ground state of Pb$^{2+}$.

In Table \ref{term-e1m1}, we have listed the term-wise contributions from 
FSRCC for selected E1 and M1 matrix elements. As expected, the DF is the
leading order (LO) term for both the matrix elements. It contributes 
$\approx 109$ and 104\% of the total value for $\langle^1S_0||D||^1P^o_1\rangle$ 
and $\langle^3P^o_1||M1||^3P^o_0\rangle$, respectively. The next leading 
order (NLO) contribution of the two matrix elements show different trends.
For $\langle^1S_0||D||^1P^o_1\rangle$, the NLO contribution of opposite 
phase of $\approx$ -20\% arises from the {\em one-valence} sector.
Whereas, for $\langle^3P^o_1||M1||^3P^o_0\rangle$, the NLO contribution is 
from the {\em two-valence} sector, the ${R_{2}}\hat O {R_{2}}$ term
gives a contribution of $\approx -3$\%. As the next dominant contribution, 
the term $\hat O {R_{2}} + {\rm h. c.}$ contributes $\approx 5$ and 1.5\%, 
respectively, to $\langle^1S_0||D||^1P^o_1\rangle$ and 
$\langle^3P^o_1||M1||^3P^o_0\rangle$ matrix elements.

For the contributions from the Breit and QED corrections to matrix elements, 
the largest contribution is observed in the case of 
$\langle^1S_0||M1||^3D_1\rangle$ transition.
The Breit contributes $\approx$ 13.0\%, whereas the contribution from the QED 
is $\approx$ 5.0\% of the total value.
transition. The largest contribution from the perturbative triples is 
observed to be $\approx$ 2.4\%, in the case of $\langle^3D_1||D||^3P^o_0\rangle$. 
Combining these two, the largest consolidated contribution from 
Breit+QED+perturbative triples is $\approx$ 20\%. Considering the high 
accuracies associated with atomic clocks, this is a significant contribution. 
Hence, it is important to include these to obtain reliable clock properties 
from theoretical calculations.

To understand the nature of electron correlations subsumed in computations
of $\alpha$, we have listed the termwise contributions from PRCC theory in 
Table \ref{pol_tw}. As evident from the table, the LO term, 
${{\mathbf T}_{1}^{(1)\dagger}{\mathbf D}}$ + h. c., contributes 
$\approx 118$\% of the total value. This is expected, as it includes the DF 
and dominant contribution from {\em core-polarization}.
For a better illustration, in Fig. \ref{fig_tbq}(d), we have shown the 
five dominant contributions from core orbitals. As discernible from the 
figure, $\approx 87$\% of the LO contribution arises from the $6s$ 
valence-electrons through the dipolar mixing with the $6p$ states. This is 
due to the larger radial extent of the $6s$ orbital. In the remaining LO
contribution,  a contribution of $\approx 8.5$\% is from the $5d$ 
core-electrons, through the dipolar mixing with $6p$ and $4f$-electrons. The 
NLO term is ${\mathbf T_{1}}{^{(1)\dagger}}{\mathbf D}T_{2}^{(0)}$, it 
contributes $\approx$ 8\%. It is to be noted that, it accounts for the dominant
{\em pair-correlation} effects through the $T_{2}^{(0)}$ operator. 
The next dominant contribution of $\approx 3$\%, which also include some part 
of {\em pair-correlation}, is from 
${\mathbf T_{2}}{^{(1)\dagger}}{\mathbf D}T_{2}^{(0)}$.

The perturbative triples and Breit each contributes $\approx 0.04$\% to 
$\alpha$. The contribution from QED is, however, significant, 
$\approx 0.7$\%. So, the cumulative contribution from Breit+QED+perturbative 
triples is $\approx$ 0.8\%. The contribution from the higher symmetry orbitals 
is estimated to be $\approx 0.34$\% of the total value.

%%%%%%%%%%%%%%%%%%%%%%%%%%%%%%%%%%%%%%%%%%%%%%%%%%%%%%%%%%%%%%%%%%%%%%%%%%%%%%
%%%%                  Section:Theoretical Uncertainty                     %%%%
%%%%%%%%%%%%%%%%%%%%%%%%%%%%%%%%%%%%%%%%%%%%%%%%%%%%%%%%%%%%%%%%%%%%%%%%%%%%%%
\section{Theoretical Uncertainty}

The theoretical uncertainty in the computed $\tau$ depends on the 
uncertainties in the E1 and M1 matrix elements and the energy denominators,
as they contribute in Eq. (\ref{e1m1}). As the experimental results are not 
available for all the E1 and M1 reduced matrix elements,
we have identified four different sources which can contribute to the 
uncertainty in E1 and M1 matrix elements. The first source of uncertainty is due 
to the truncation of the basis set in our calculation. As discussed in the basis 
convergence section, our calculated values of E1 and M1 reduced matrix elements 
converge to the order of $10^{-3}$ or smaller with basis. Since this is a very 
small change, we may neglect this uncertainty. The second source of 
uncertainty arises from the truncation of the dressed 
Hamiltonian $\tilde{H}^{\rm e}_{\rm hfs}$ at the second order of $T^{(0)}$ in 
the properties calculation. In our earlier work \cite{mani-10}, using an 
iterative scheme, we found that the terms with third and higher orders in 
$T^{(0)}$ contribute less than 0.1\%. So, we consider 0.1\% as an upper bound 
for this source. The third source is due to the partial inclusion of triple 
excitations in the properties calculation. Since the perturbative triples 
account for the leading order terms of triple excitation, the contribution 
from remaining terms will be small. Based on the analysis from our previous 
works \cite{ravi-20, chattopadhyay-15} we estimate the upper bound from this 
source as 0.72\%. The fourth source of uncertainty could be associated with the 
frequency-dependent Breit interaction which is not included in the present 
calculations. However, in our previous work \cite{chattopadhyay-14}, using a 
series of computations using GRASP2K we estimated an upper bound on this 
uncertainty as $0.13$\% in Ra. So, for the present work, we take $0.13$\% as 
an upper bound from this source. There could be other sources of theoretical 
uncertainty, such as the higher order coupled perturbation of vacuum 
polarization and self-energy terms, quadruply excited cluster operators, etc. 
However, in general, these all have much lower contributions to the properties 
and their cumulative theoretical uncertainty could be below 0.1\%.
Uncertainty in the energy denominator is estimated using the relative 
errors in the energy difference of $^3P^{o}_1$, $^1P^{o}_1$, $^3S_1$ 
and $^3D_1$ intermediate states with respect to $^3P^{o}_0$. Among all the 
intermediate states, $^1P^{o}_1$ and $^3D_1$ states contribute dominantly, 
through routes 1 and 2, respectively, to the lifetime. The relative 
errors in the energy difference of these states with $^3P^{o}_0$ are 1.27\% 
and 0.78\%, respectively. Since they correspond to the dominant contributions, 
we have taken them as the uncertainty in the energy denominator.  
%The other source of theoretical uncertainty which will contribute to the 
%lifetime of clock state is the QED corrections at the level of Eq. (\ref{e1m1}). 
%To estimate this uncertainty, we refer to the Refs. \cite{shabaev-05a, shabaev-05b}. 
%In these works Shabaev and collaborators have computed the one-loop QED corrections 
%to the properties of Cs and Fr. The reported total contribution is about 0.28\%. So, 
%based on these works we consider 0.3\% as the upper bound from this source of 
%uncertainty. 
By combining the upper bounds of all the uncertainties, the theoretical 
uncertainty associated with the lifetime of the clock state is $\approx$ 4.8\%. 
It should, however, be noted that the uncertainty in the value of $\alpha$ is 
much smaller, about 1.5\% \cite{ravi-22}

%%%%%%%%%%%%%%%%%%%%%%%%%%%%%%%%%%%%%%%%%%%%%%%%%%%%%%%%%%%%%%%%%%%%%%%%%%%%%%%
%%%%                        Conclusions                                    %%%%
%%%%%%%%%%%%%%%%%%%%%%%%%%%%%%%%%%%%%%%%%%%%%%%%%%%%%%%%%%%%%%%%%%%%%%%%%%%%%%%
\section{Conclusions}

We have employed an all-particle multireference Fock-space relativistic 
coupled-cluster theory to examine the clock transition properties in 
Pb$^{2+}$. We have computed the excitation energies of several 
low lying states, and the E1 and M1 transition amplitudes for all the allowed 
transitions within the model space considered. These are then used to 
calculate the lifetime of the clock state. Moreover, using PRCC 
theory, we also calculated the electric dipole polarizability for the ground 
state of Pb$^{2+}$. In all these calculations, to obtain 
accurate properties results, we incorporate the corrections from the 
relativistic and QED effects. The dominant contribution from triple 
excitations is incorporated though perturbative triples and a fairly large 
basis sets is used to achieve the convergence of the properties.

Our computed excitation energies are in excellent agreement with experimental
values for all the states. Our result of $\tau$ is about 8.5\% larger than the 
previous result obtained using CI+MBPT \cite{beloy-21}. The reason for the 
higher $\tau$ in the present calculation could partially be attributed to the 
better inclusion of {\em core-core} and {\em core-valence} electron 
correlations in the FSRCC theory. In addition, to account for the 
{\em valence-valence} correlation more accurately, we also incorporated 
the contributions from the higher energy configurations $6s6d$ and $6s7s$ in our
calculation. Based on our analysis, we find that this contributes 
$\approx 3.3$\% of the total lifetime. In addition, from our study, we find 
that the contributions from the perturbative triples and Breit+QED corrections 
are crucial to get reliable $\tau$. They are observed to contribute 
$\approx$ -2.2\% and 1.1\%, respectively. Our recommended value of dipole 
polarizability is in good agreement with the available experimental value, 
with a small difference of $\approx 3$\%. Based on our analysis on theoretical 
uncertainty, the upper bound on uncertainty for calculated lifetime is 
$\approx 4.8$\%, whereas for polarizability it is $\approx 1.5$\%.

%%%%%%%%%%%%%%%%%%%%%%%%%%%%%%%%%%%%%%%%%%%%%%%%%%%%%%%%%%%%%%%%%%%%%%%%%%%%%%%
%%%%%%%%%%%%%%%%%%%%%%%%%%%%%%%%%%%%%%%%%%%%%%%%%%%%%%%%%%%%%%%%%%%%%%%%%%%%%%%
\begin{acknowledgments}
The authors wish to thank Suraj Pandey for the useful discussion. Palki
acknowledges the fellowship support from UGC (BININ04154142), Govt. of India. B. K. M
acknowledges the funding support from SERB, DST (CRG/2022/003845). Results
presented in the paper are based on the computations using the High
Performance Computing clusters Padum at IIT Delhi and PARAM HIMALAYA 
facility at IIT Mandi under the National Supercomputing Mission of Government of India.
\end{acknowledgments}

%%%%%%%%%%%%%%%%%%%%%%%%%%%%%%%%%%%%%%%%%%%%%%%%%%%%%%%%%%%%%%%%%%%
%%%%%%%               Appendix                          %%%%%%%%
%%%%%%%%%%%%%%%%%%%%%%%%%%%%%%%%%%%%%%%%%%%%%%%%%%%%%%%%%%%%%%%%%%%
\appendix

\section{Convergence of $\alpha$ and E1 and M1 matrix elements}

In Table \ref{basis_conv}, we provide the trend of the convergence of 
dipole polarizability and E1 and M1 matrix elements as a function of the 
basis size. As it is evident from the table, all the properties converge to 
the order of $10^{-3}$ or less in the respective units of the properties.

%%%%%%%%%%%%%%%%%%%%%%% polarizability
\begin{table}
  \caption{Convergence trend of $\alpha$, and E1 and M1 matrix elements 
	as a function of basis size.} 
  \label{basis_conv}
  \begin{ruledtabular}
  \begin{tabular}{cccc}
	  Basis & $\alpha$ & $\langle^1S_0|| D ||^1P^o_1\rangle$ & $\langle^3P^o_1|| M1 ||^3P^o_0\rangle$  \\
       \hline
	  $ 96: 20s15p12d7f4g0h$  & $15.919$  & 2.0512  & 1.3158 \\   
	  $103: 19s16p13d8f3g2h$  & $15.660$  & 2.0268  & 1.3153  \\         
	  $114: 20s17p14d9f4g3h$  & $14.979$  & 1.9966 & 1.3198  \\       
	  $125: 21s18p15d10f5g4h$ & $14.467$  & 1.9873 & 1.3134  \\  
	  $136: 22s19p16d11f6g5h$ & $14.274$  & 1.9888 & 1.3126  \\   
	  $147: 23s20p17d12f7g6h$ & $14.205$  & 1.9883 & 1.3121  \\  
	  $158: 24s21p18d13f8g7h$ & $14.175$  & 1.9875 & 1.3118  \\             
	  $169: 25s22p19d14f9g8h$ & $14.173$  & 1.9854 & 1.3117  \\  
 \end{tabular}                                         
 \end{ruledtabular}
\end{table}            
%%%%%%%%%%%%%%

%%%%%%%%%%%%%%%%%%%%%%%%%%%%%%%%%%%%%%%%%%%%%%%%%%%%%%%%%%%%%%%%%%%%%%%%%%%%%%%%%%%%%%%%%%
%%%%%%%%%%%%%%%%%%%%           References        %%%%%%%%%%%%%%%%%%%%%%%%%%%%%%%%%%%%%%%%%
%%%%%%%%%%%%%%%%%%%%%%%%%%%%%%%%%%%%%%%%%%%%%%%%%%%%%%%%%%%%%%%%%%%%%%%%%%%%%%%%%%%%%%%%%%
\bibliography{reference}
\end{document}